\documentclass[twocolumn,superscriptaddress,showpacs,preprintnumbers,nofootinbib,amsmath,amssymb,floatfix,aps]{revtex4-1}
\usepackage{dcolumn}
\usepackage{bm}
\usepackage{color}

\usepackage{graphicx}
\usepackage{amsmath}
\usepackage{amssymb}

\def\lsim{\buildrel < \over {_{\sim}}}
\def\gsim{\buildrel > \over {_{\sim}}}

\def\beq{\begin{equation}}
\def\eeq{\end{equation}}
\def\be{\begin{eqnarray}}
\def\ee{\end{eqnarray}}

\begin{document}
\title{Contribution of two particle-two hole final states to the nuclear response}
\author{Omar Benhar}
\affiliation{INFN and Department of Physics, ``Sapienza'' University, I-00185 Roma, Italy}
\affiliation{Center for Neutrino Physics, Virginia Tech, Blacksburg, Virginia 24061, USA} 
\author{Alesssandro Lovato}
\affiliation{Argonne Leadership Computing Facility, Argonne National Laboratory, Argonne, Illinois 60439, USA}
\affiliation{Physics Division, Argonne National Laboratory, Argonne, Illinois 60439, USA}
\author{Noemi Rocco}
\affiliation{INFN and Department of Physics, ``Sapienza'' University, I-00185 Roma, Italy}

\date{\today}
\begin{abstract}
The excitation of two particle-two hole final states in neutrino-nucleus scattering
has been advocated by many authors as the source of the excess cross section observed by the MiniBooNE Collaboration 
in the quasi elastic sector. We analyse the mechanisms leading to the appearance of these final states, and illustrate their
significance through the results of accurate calculations of the nuclear electromagnetic response in the transverse channel.
A  novel approach, allowing for a consistent treatment of the amplitudes involving one- and two-nucleon currents in the 
kinematical region in which the non relativistic approximation breaks down is outlined, and its preliminary results are reported.
\end{abstract}
\maketitle

\section{Introduction}

Experimental studies of neutrino-nucleus interactions  
 carried out over the past decade \cite{K2K,MiniBooNE_Q2dist,nomad,MiniBooNE_d2sigma} have
 provided ample evidence of the inadequacy of the Relativistic Fermi Gas Model (RFGM), routinely 
 employed in event generators, to account for both the complexity of nuclear dynamics and the variety 
of reaction mechanisms---other than single nucleon knock out---contributing to the observed cross section.

A striking manifestation of the above problem is the large discrepancy between the predictions of Monte
 Carlo simulations and the double differential charged current (CC) quasi elastic (QE) cross section 
 measured by the MiniBooNE Collaboration using a carbon target  \cite{MiniBooNE_d2sigma}. 

As pointed out by the authors of Ref. \cite{coletti}, improving the treatment of nuclear effects, 
which turns out to be  one of the main sources of systematic uncertainty in the oscillation analysis \cite{T2K}, 
will require the development of a {\em comprehensive} and {\em consistent} description of neutrino-nucleus 
 interactions, {\em validated} through extensive comparison to the large body of electron-nucleus 
 scattering data \cite{PRD,RMP}.
  
The main difficulty involved in the generalisation of the approaches successfully employed to study electron scattering 
to the case of neutrino interactions stems from the fact that, while 
the energy of the electron beam is fixed, in neutrino scattering the measured cross section results from the  average over
different beam energies, broadly distributed according to a flux $\Phi$.  Therefore, a measurement of the energy of the
outgoing charged lepton in a CC QE interaction {\em does not} specify the energy transfer to the nuclear target,  
which largely determines the reaction mechanism.  As shown in Refs. \cite{benhar:neutrino2010,benhar:nufact}, the 
MiniBooNE double differential cross section corresponding to a specific muon energy bin turns out to receive comparable 
contributions from different mechanisms, which must be all taken into account in a consistent fashion.

Many authors have suggested that the excess CC QE cross section observed by the MiniBooNE collaboration is to be ascribed
to the occurrence of events with two particle-two hole final states, not taken into account by the RFGM employed for 
data analysis  \cite{martini,coletti, nieves}. 
The description of these processes
within a realistic model of nuclear dynamics requires that all mechanisms leading to their occurrence---Initial State Correlations (ISC) among 
nucleons in the target nucleus, Final State Correlations (FSC) between the struck nucleon and the spectator particles, and interactions 
involving two-nucleon meson-exchange currents (MEC)---be included. Within the Independent Particle Model (IPM) of the nucleus,  however, 
correlations are not taken into account, and two particle-two hole final states can only be excited through the action of two-body operators, such as those involved 
in the definition of MEC.

In this paper, we analyse the mechanisms leading to the appearance of two particle-two hole final states in the response of interacting many-body systems,  and 
argue that the interference between amplitudes involving one- and two-nucleon currents plays an important role. 
This feature clearly emerges from the results of a calculation of the transverse electromagnetic response of $^{4}$He and of the corresponding sum 
rule of $^{12}$C, 
computed using state-of-the-art models of the nuclear hamiltonian and
currents, within the Green's Function Monte Carlo (GFMC) computational scheme  \cite{Carlson:2002}.   

In view of the extension of our study to the kinematical regime in which the non relativistic approximation 
is no longer applicable, we also outline a novel approach, based on a generalisation of the factorisation {\em ansatz}, underlying 
the spectral function formalism. This scheme,  allowing for a consistent treatment  of one- and two-nucleon 
current contributions, appears to be quite promising for applications to neutrino scattering.
 
The structure of the nuclear cross section, as well as its expression in terms of longitudinal and transverse structure functions, are
reviewed in Section~\ref{response}, while Section~\ref{NMBT} describes the theoretical approaches, based on nuclear many-body theory,  
developed to study electron-nucleus scattering. In Section~\ref{nonrel} we discuss the non relativistic regime and the results of GFMC 
calculations, while in Sections~\ref{rel} and \ref{rel2} we derive the explicit expression of the two particle-two hole contribution to the cross section
obtained from our approach, and discuss the preliminary results of its application. Finally, in Section~\ref{conclusions} we summarise our findings and state the conclusions.

\section{Nuclear cross section and response functions}
\label{response}

In the one-photon-exchange approximation, the  double differential electron-nucleus cross section can be written
in the form
\beq
\label{xsec}
\frac{d^2\sigma}{d E_{e^\prime} d\Omega_{e^\prime}}=\frac{\alpha^2}{q^4}\frac{E_{e^\prime}}{E_e}L_{\mu\nu}W_A^{\mu\nu} \ ,
\eeq
where $k_e=(E_e,{\bf k}_e)$ and $k_{e^\prime}=(E_{e^\prime},{\bf k}_{e^\prime})$ are the four-momenta of the incoming and outgoing electrons, respectively, $\alpha = 1/137$ is the fine structure constant, $d\Omega_{e^\prime}$ is the differential solid angle in the direction specified by ${\bf k}_{e^\prime}$, and $q=k_e - k_{e^\prime} =(\omega,{\bf q})$ is the four momentum transfer.

The lepton tensor $L_{\mu\nu}$ is  completely determined by lepton kinematics, while the nuclear response is described by the tensor $W_A^{\mu\nu}$, defined as
\begin{align}
\label{response:tensor}
W^{\mu \nu}_A({\bf q},\omega) =\sum_N \langle  0| J_A^\mu | N \rangle  \langle  N | J_A^\nu |   0 \rangle \delta^{(4)}(P_0+q-P_N)   \ ,
\end{align}
where $| 0 \rangle$ and $| N \rangle$ denote the initial and final hadronic states, the four-momenta of which are $P_0\equiv ( E_0,{\bf p}_0 )$ and 
$P_N \equiv (E_N,{\bf p}_N) $.
 The nuclear current can be written as a sum of one- and two-nucleon contributions, according to (see, e.g., Ref. \cite{2NC})
\beq
\label{nuclear:current}
J_A^\mu= \sum_i j^\mu_i+\sum_{j>i} j^\mu_{ij} \ .
\eeq
The current $j_i^\mu$ describes interactions involving a single nucleon. In the QE sector,  it can be expressed in terms of the measured vector  
form factors \cite{VFF}. The two-nucleon contribution $j^\mu_{ij}$, on the other hand, accounts for processes 
in which the beam particle couples to the currents arising from meson exchange between two interacting nucleons.

Equation \eqref{xsec} can be rewritten in terms of two response functions, denoted $R_L({\bf q}, \omega)$ 
and $R_T({\bf q},\omega)$, describing interactions with longitudinally (L) and transversely (T) polarised photons, 
respectively. The resulting expression reads 
\begin{align}
\frac{d^2\sigma}{d E_e^\prime d\Omega_e} &  =\left( \frac{d \sigma}{d \Omega_e} \right)_{\rm{M}} \Big[  A_L(|{\bf q}|,\omega,\theta_e)  R_L(|{\bf q}|,\omega) \\
& + A_T(|{\bf q}|,\omega,\theta_e)  R_T(|{\bf q}|,\omega) \Big] \ ,
\end{align}
where 
\begin{align}
A_L = \Big( \frac{q^2}{{\bf q}^2}\Big)^2  \ \ \ , \ \ \ A_T = -\frac{1}{2}\frac{q^2}{{\bf q}^2}+\tan^2\frac{\theta_e}{2}  \ , 
\end{align}
and $( d \sigma/d \Omega_e)_{\rm{M}}= [ \alpha \cos(\theta_e/2)/4 E_e\sin^2(\theta_e/2) ]^2$ is the Mott cross section.

The $L$ and $T$ structure functions can be readily expressed in terms of the components of the response tensor of Eq. \eqref{response:tensor}. Choosing the 
$z$-axis along the direction of the momentum transfer one finds
\begin{align}
\label{RL}
R_L & = W_A^{00} \\
\label{RT}
R_T &= \sum_{ij=1}^3\Big(\delta_{ij}-\frac{q_iq_j}{{\bf q}^2}\Big)W^{ij}_A = W^{xx}_A +  W^{yy}_A \ . 
\end{align}

Note that  the above expressions are completely general, and describe processes involving both one- and two-nucleon current operators.\\

It follows from Eqs. \eqref{xsec} and \eqref{response:tensor} that the nuclear cross section and response functions can be written as a sum of
contributions corresponding to different hadronic final states $|N \rangle$. Consider, for example, the case of QE scattering, in which the final 
state particles are nucleons only.  For a carbon target we find 
\begin{align}
\label{carbon:states}
|N\rangle= |^{11}B, p \rangle\ ,\ |^{11}C, n \rangle\ , |^{10}B, pn \rangle\ ,|^{10}Be, pp \rangle \dots,  
\end{align}
where the residual nucleus can be in any {\em bound} state. 

The states $|N \rangle$ are usually classified according to the number of nucleons excited to the 
continuum, and referred to as one particle-one hole (1p1h), two particle-two hole (2p2h), etc. In Eq. \eqref{carbon:states},  $|^{11}B, p \rangle$ and 
$|^{11}C, n \rangle$ are 1p1h states, while  $|^{10}B, pn \rangle$  and $|^{10}Be, pp \rangle$ are 2p2h states.

Neglecting the contributions of final states involving more than two nucleons in the continuum, the cross section can be written as
\beq
\label{xsec:split}
d\sigma = d\sigma_{1p1h}+d\sigma_{2p2h} \propto L_{\mu\nu} ( W^{\mu\nu}_{1p1h}+ W^{\mu\nu}_{2p2h} ) \ .
\eeq

We recall that, in scattering processes involving {\em interacting} many-body systems, 2p2h final states 
can be produced through the action of {\em both} one- and two-nucleon currents \footnote{It should be kept in mind  that 1p1h final states can also be 
excited by both one- and two-nucleon currents. }.
However, in order for the matrix element of a one-body operator between the target ground state and a 2p2h final state to be non vanishing, 
the effects of dynamical nucleon-nucleon ($NN$) correlations must be included in the description of the nuclear wave functions. 

Correlations give rise to virtual scattering between nucleons in the target nucleus, leading to the excitation of the participating 
particles to continuum states. The ISC contribution to the 2p2h amplitude arises from processes in which the 
beam particle couples to {\em one} of these high-momentum nucleons.  The FSC contribution, on the other hand, originates from scattering 
processes involving the struck nucleon and one of the spectator particles, that also result in the appearance of  2p2h final states.

\section{Many-body theory of the nuclear response}
\label{NMBT}

As discussed in the previous Section, the calculation of the nuclear response requires the evaluation of the transition amplitudes $\langle  0| J_A^\mu | N \rangle$, 
involving both one- and two-nucleon current operators,  as well as all possible final states. 
The initial state can be accurately described within the framework of non relativistic many-body theory using realistic models of the nuclear hamiltonian,
strongly constrained by nucleon-nucleon scattering data and nuclear phenomenology. 
The final state and the current operator, on the other hand, depend on momentum transfer, and their calculation in the kinematical region in which the
non relativistic picture breaks down necessarily implies additional assumptions.

\subsection{Non relativistic regime}
\label{nonrel}

The approach based on the GFMC computational scheme provides a suitable framework to carry out accurate calculations of a variety of nuclear properties 
in the non relativistic regime, typically corresponding to $|{\bf q}| \lsim 500 \ {\rm MeV}$ (for a recent review of Quantum Monte Carlo methods for nuclear physics see, e.g., Ref. \cite{QMCreview}). 

Valuable information on the L and T responses can be obtained from their Laplace transforms, also referred to as Euclidean responses, defined as 
\beq
\widetilde{E}_{T,L}({\bf q}, \tau)= \int_{\omega_{\rm{el}}}^\infty \,{d\omega} e^{-\omega \tau}R_{T,L}({\bf q}, \omega)\ .
\eeq
The lower integration limit $\omega_{\rm{el}}= {\bf q}^2/2M_A$, $M_A$ being the mass of the target nucleus, is the threshold of elastic scattering---corresponding to the 
$|N \rangle = |0 \rangle$ term in the sum of Eq. \eqref{response:tensor}---the contribution of which  is excluded.

Within GFMC, the Euclidean responses are evaluated from 
\begin{align}
\nonumber
\widetilde{E}_L({\bf q},\tau) & = \langle 0| \rho^\ast({\bf q}) e^{-(H-E_0)\tau}  \rho({\bf q})|0\rangle \\ 
& -  |\langle 0 | \rho({\bf q}) | 0 \rangle|^2 e^{-\omega_{\rm el} \tau} \ ,
\label{eq:eucL_mat_el}
\end{align}
and 
\begin{align}
\nonumber
\widetilde{E}_T({\bf q},\tau) & = \langle 0| {\bf j}_T^\dagger({\bf q}) e^{-(H-E_0)\tau} {\bf j}_T({\bf q})|0\rangle \\ 
& -  |\langle 0 | {\bf j}_T({\bf q}) | 0 \rangle|^2 e^{-\omega_{\rm el} \tau} \ ,
\label{eq:eucT_mat_el}
\end{align}
where $\rho({\bf q})$ and ${\bf j}_T({\bf q})$ denote non relativistic reductions of the nuclear charge and transverse current operators, respectively \cite{Carlson:2002}.

Note that, although the states $|N \rangle \neq | 0 \rangle$ do not appear explicitly in Eqs. \eqref{eq:eucL_mat_el} and \eqref{eq:eucT_mat_el}, the Euclidean responses  
include the effects of final state interactions (FSI) of the particles involved in the electromagnetic interaction, both among themselves and with the spectator nucleons. 

The Euclidean responses at $\tau=0$ are directly related to the sum rules of the L and T responses, obtained from $\omega$-integration after removing the 
trivial energy and momentum dependence associated with the nucleon form factor \cite{lovato12C}:
\begin{align} 
\label{sumrules}
S_{T,L}(|{\bf q}|) = \frac{ C_{T,L}}{[G_E^p(Q^2_{\rm QE})]^2}  \int_{\omega_{\rm{el}}}^\infty \,{d\omega} R_{T,L}({\bf q}, \omega)\  .
\end{align}
In the above equation, $G_E^p(Q^2_{\rm QE})$ is the electric proton form factor evaluated in quasi elastic kinematics, i.e. at $Q^2_{\rm QE} = {\bf q}^2 - \omega^2_{\rm QE}$, with 
$\omega_{\rm QE} =  ( \sqrt{ {\bf q}^2 + m^2} - m)$, where $m$ is the proton mass. The coefficients appearing in Eq.~\eqref{sumrules} are defined as
\begin{align}
 C_L = \frac{1}{Z}  \ \ \ \ \ , \ \ \ \ \ C_T = \frac{2}{Z \mu^2_p + N \mu^2_n} \ \frac{m^2}{{\bf q}^2} \ ,
\end{align}
where $Z$ is the proton charge, $N = A -Z$ is the number of neutrons and $\mu_p$ and $\mu_n$ are the proton and neutron magnetic moments, respectively. 

The inversion of the Laplace transform, needed to retrieve the energy dependence of the responses, is long known to involve severe difficulties.  
A groundbreaking result has been recently reported by the authors of Ref.~\cite{inversion}, who exploited the maximum entropy technique to obtain 
the L and T responses of $^4$He. 

\begin{figure}[h!]
 \includegraphics[scale= 0.5]{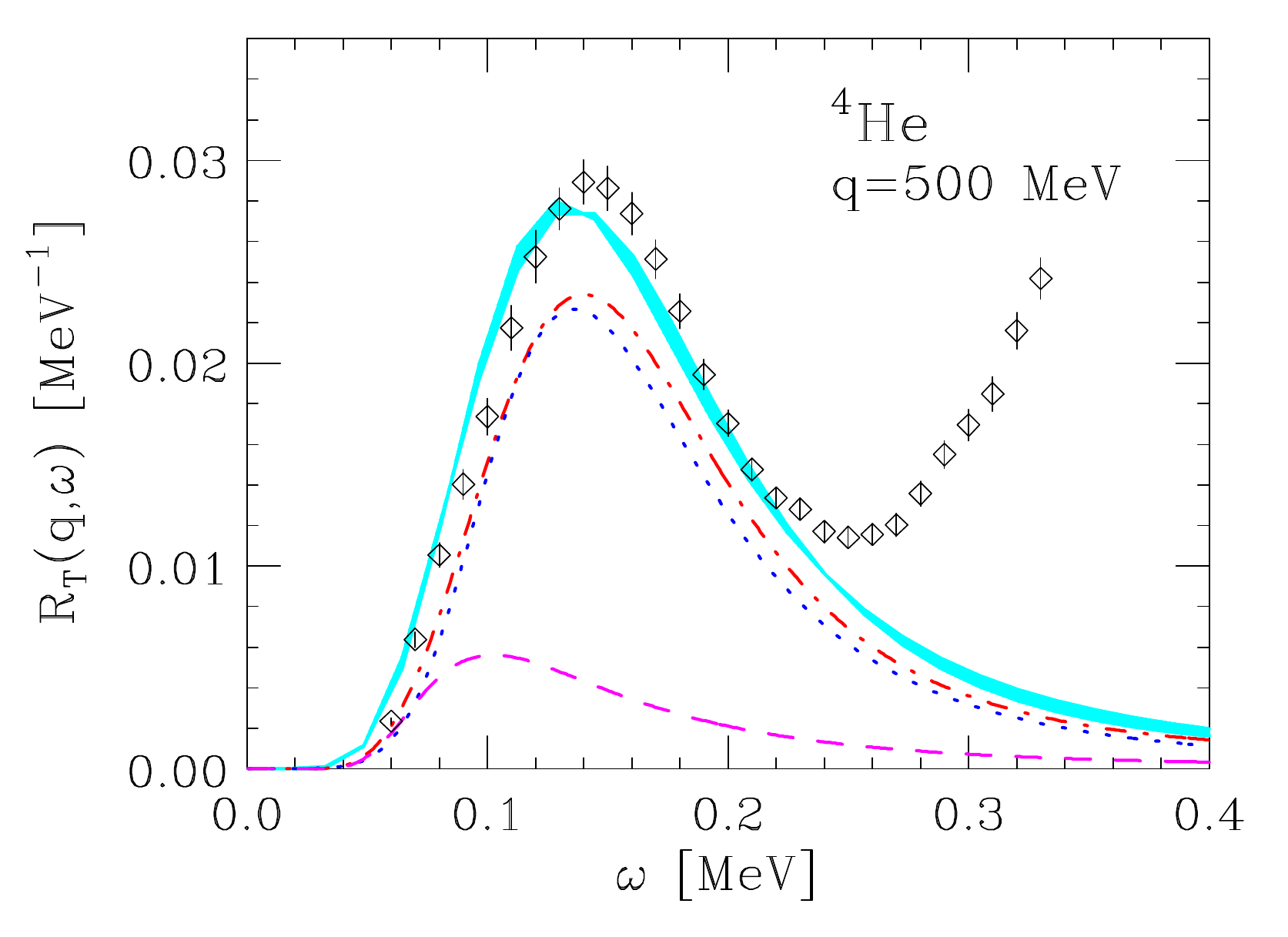} 
\vspace*{-.2in}
\caption{(color online) Transverse response function of $^4$He, obtained within the approach of Ref.~\cite{inversion}. The shaded area shows the results of the full 
calculation, with the associated uncertainty arising from the inversion of the Euclidean response, while the dotted line 
has been obtained including the one-nucleon current only. The dot-dash line represents the response computed 
neglecting the interference term, the contribution of which is displayed by the dashes. The data are taken from Ref.~\cite{Carlson:2002}.}
\label{4He}
\end{figure}

Figure \ref{4He} shows the breakdown of the transverse response of $^4$He at $|{\bf q}| = 500$ MeV---computed within the approach of Ref.~\cite{inversion}---into one-nucleon current, 
two-nucleon current and interference contributions. Note that the quantity displayed in the figure is   
normalised dividing by the squared proton form factor. 

It clearly appears that including the two-nucleon currents leads to a sizeable enhancement of the response, and that the large positive contribution of the interference term peaks at energy loss $\omega < \omega_{\rm QE}$. 
This feature is a direct consequence of nucleon-nucleon correlations, neglected in the mean field approach.  The agreement between the GFMC results and the data of Ref.~\cite{Carlson:2002} 
turns out to be remarkably good. 

The extension of the procedure employed to obtain the $^4$He response to heavier nuclei, such as carbon, is still out of reach of the available computational 
capabilities. However, valuable information can be extracted from the analyses of the sum rules. 

\begin{figure}[h!]
 \includegraphics[scale= 0.5]{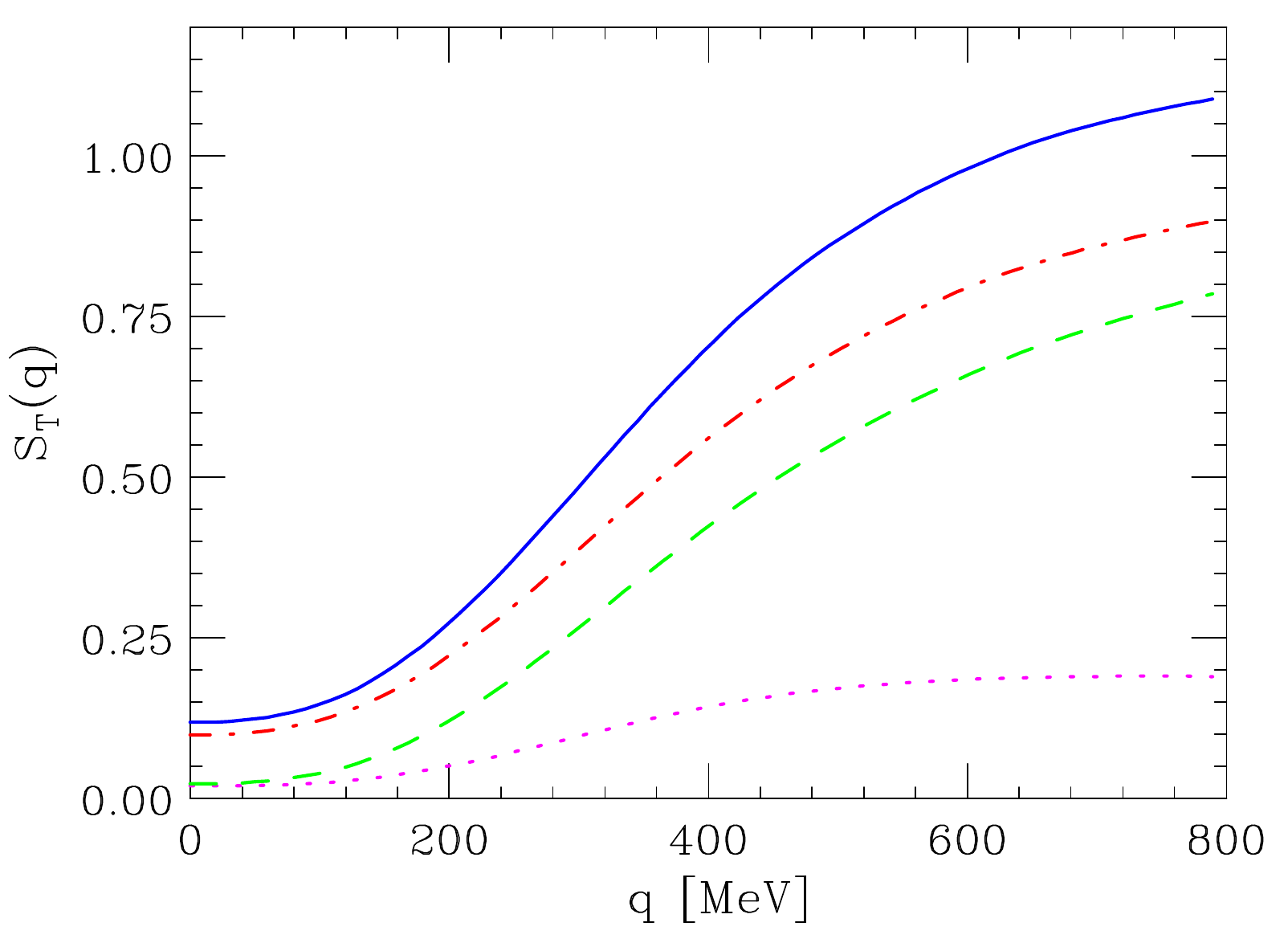} 
\vspace*{-.2in}
\caption{(color online) Sum rule of the electromagnetic response of carbon in the transverse channel. The dashed line 
shows the results obtained including the one-nucleon current only, while the solid line corresponds to the 
 full calculation. The dot-dash line represents the sum rule computed neglecting the interference 
term, the contribution of which is displayed by the dotted line. The results are normalised so that the dashed line approaches 
unity as $|{\bf q}| \to \infty$. Monte Carlo errors bars are not visible on the scale of the figure.}
\label{sumrule}
\end{figure}

The results of numerical calculations of the carbon $S_T(|{\bf q}|)$, displayed in Fig.~\ref{sumrule}, clearly show that interference terms provide a sizeable fraction of the sum rule.
At momentum transfer $|{\bf q}| \gsim 300$ MeV, their contribution turns out to be comparable to---in fact even larger than---that obtained squaring 
the matrix element of the two-nucleon current.

\subsection{Relativistic regime: the factorisation {\em ansatz}}
\label{rel}

The results of Figs.~\ref{4He} and \ref{sumrule} clearly point to the need for a consistent treatment of correlations and MEC, within a 
formalism suitable for application in the kinematical region in which the non relativistic approximation is known to fail. This section
describes the derivation of  the approach based on factorisation of the nuclear matrix elements. For ease of presentation, we will consider
the response of uniform and isospin symmetric nuclear matter. However, the generalisation to atomic nuclei does not involve any 
substantial problems.
 
The effects of ISC on the nuclear cross section at large momentum transfer can be taken into account using the 
spectral function formalism \cite{PKE,LDA}. The conceptual framework underlying this approach 
is provided by the impulse approximation (IA), i.e. the assumption that at momentum transfer such that
$|{\bf q}|^{-1} \ll d$, $d$ being the average separation distance between nucleons in the target nucleus,  
the  nuclear cross section reduces to the incoherent sum of cross sections describing scattering processes involving  individual nucleons. As a consequence, the contribution of 
the two-nucleon current can be disregarded, and the final state  $| N \rangle$ of Eq.~\eqref{response:tensor} can be written 
in the factorized form
\beq
\label{fact1}
| N \rangle = | {\bf p}\rangle \otimes | n_{A-1}, {\bf p}_n \rangle \ .
\eeq
In the above equation,  $|{\bf p}\rangle$ is the state of a non interacting nucleon carrying momentum ${\bf p}$, while  $| n_{A-1}, {\bf p}_n \rangle$ describes the 
 $(A-1)$-particle spectator system in the state $n$, with momentum ${\bf p}_n$. Note that, owing to $NN$ correlations,  $| n_{A-1}, {\bf p}_n \rangle$ is not restricted to be  
a bound  state [see Eq.~\eqref{carbon:states}].

Within the IA, the contribution to the nuclear cross section arising from interactions involving the one-nucleon current 
can be written in terms of the cross sections of elementary scattering processes off individual nucleons, the momentum (${\bf k}$) and removal energy ($E$) of which are distributed 
according to the spectral function $P({\bf k},E)$ \cite{PKE}, defined as
\beq
\label{pke1}
P({\bf k},E) = \sum_n |\langle n_{A-1}, {\bf p}_n | a_{\bf k} | 0 \rangle |^2 \delta(E+E_0-E_n) \ .
\eeq
In the above equation,  $E_n$ is the energy of the $(A-1)$-nucleon state, and the operator $a_{\bf k}$ removes a nucleon of momentum ${\bf k}$ from the nuclear ground state. 

The resulting expression of the cross section is \cite{RMP} 
\beq
\label{sigma1}
d\sigma_{IA} =  \ \sum_i \int \,d^3k \ dE \   P({\bf k},E) \ d\sigma_{i} \   . 
\eeq
Note that $P({\bf k},E)$ describes an intrinsic property of the target nucleus, independent of momentum transfer, and as such can be safely obtained from non relativistic
many-body theory. On the other hand, the matrix elements of the nucleon current entering the definition of  $d\sigma_{i}$ can be computed using its fully relativistic form. 

Exploiting  the K\"all\'en-Lehman representation of the two-point Green's function,
the spectral function appearing in  Eq.~\eqref{sigma1},  can be conveniently split
into two parts, displaying  distinctly different energy dependences \cite{BFF_split}. The single particle part $P_{1h}({\bf k},E)$, obtained from Eq.~\eqref{pke1} including 
{\em bound} 1h states only, exhibits a pole at $E=-e_k$, $e_k$ being the energy of a nucleon in the hole state of momentum ${\bf k}$. The  continuum part, on the 
other hand, is smooth, and extends to large values of energy and momentum. Its leading term, corresponding to 2h1p states of 
the residual $(A-1)$-particle system in which one nucleon is excited to a state outside the Fermi sea, can be written in the form
\begin{widetext}
\begin{align}
P_{2h1p} ({\bf k},E) = \int {d^3h d^3h^\prime d^3p^\prime}|\Phi^{h h^\prime p^\prime}_{k}|^2 
 \theta(k_F - |{\bf h}|) \theta(k_F - |{\bf h^\prime}|)  \theta(|{\bf p^\prime}|- k_F)   
  \delta(E + e_h + e_{h^\prime} - e_{p^\prime})  \ , 
\end{align}
\end{widetext}
where the integration includes a sum over the indices associated with discrete degrees of freedom, and
\begin{align}
\label{def:phi2h1p}
\Phi^{h h^\prime p^\prime}_{k} = \langle 0 | \{ | {\bf k} \rangle \otimes | {\bf h} {\bf h}^\prime {\bf p}^\prime  \rangle   \} \ .
\end{align}
Note that momentum conservation requires that the expression of $\Phi^{h h^\prime p^\prime}_{k}$ involve 
a $\delta({\bf h} + {\bf h^\prime} - {\bf p^\prime} - {\bf k})$. 

As pointed out above, in the presence of ground state correlations both parts of the spectral function provide non vanishing contributions to the cross section of Eq.~\eqref{sigma1}.
 
\begin{figure}[h!]
\includegraphics[scale= 0.45]{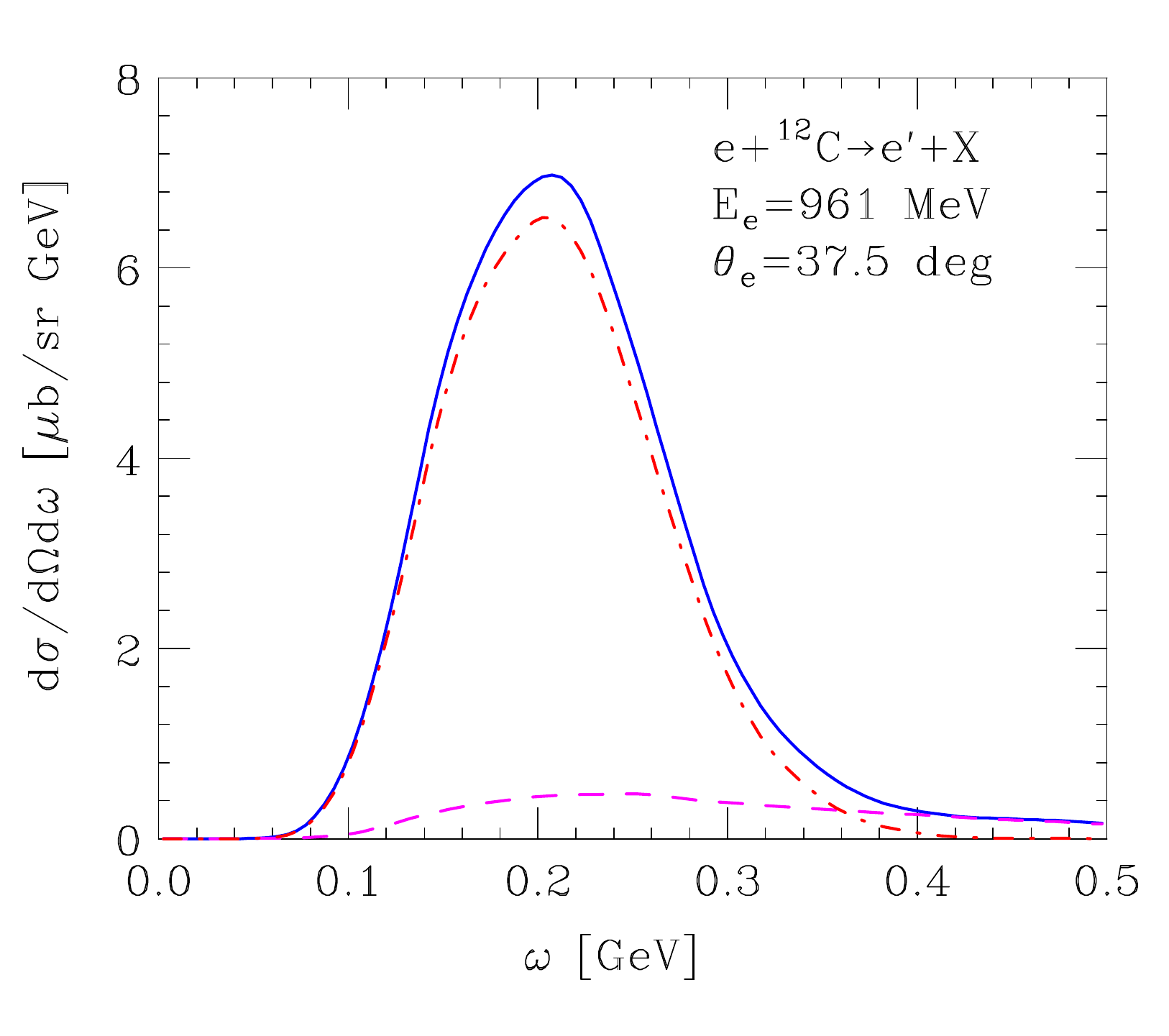} 
\vspace*{-.1in}
\caption{(color online) Cross section of the process $e+^{12}C \to e^\prime + X$ at beam energy $E_e= 961$ MeV and electron scattering angle $\theta_e= 37.5$~deg, 
computed using Eq. \eqref{sigma1} with the spectral function of Ref.~\cite{LDA}. The solid line shows the results of the full calculation, while
the breakdown into 1p1h and 2p2h contributions  is illustrated by the dot-dash and dashed  lines, respectively.}
\label{ISC}
\end{figure}

Figure \ref{ISC} shows the 1p1h and 2p2h components of the electron-carbon cross section arising from ISC. 
The calculations have been performed at $E_e = 961 \ {\rm MeV}$ and $\theta_e = 37.5 \ {\rm deg}$, using Eq. \eqref{sigma1} with the spectral function 
of Ref.  \cite{LDA} and the parametrisation of the nucleon form factors of Ref. \cite{BBBA}.
The solid line corresponds to the results of the full calculation, while the dot-dash and dashed lines have been obtained 
using the pole and continuum parts of the spectral function, which amounts to taking into account only 1p1h or 2p2h final states, respectively. The distinct energy dependence 
of the 2p2h contribution, providing $\sim 10 \%$ of the total QE cross section,   is clearly visible.
 
\begin{figure}[h!]
\includegraphics[scale=0.45]{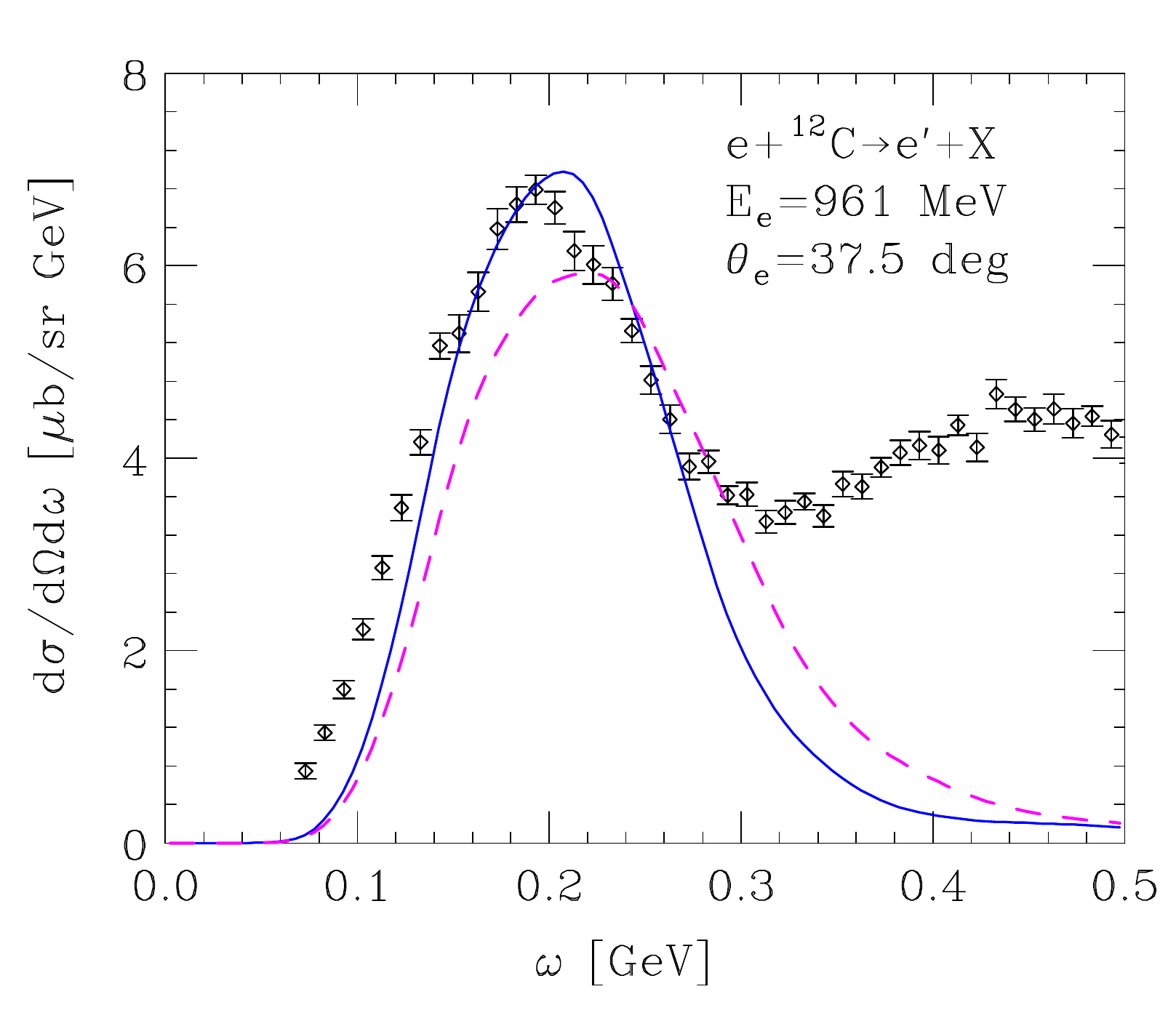}
\vspace*{-.1in}
\caption{(color online) Electron-carbon cross section obtained from Eq. \eqref{sigma1} using relativistic (solid line) and non relativistic (dashed line) kinematics.
The experimental data are from Ref. \cite{carbon:data}.}
\label{relkin}
\end{figure}

The importance of  relativistic effects can be gauged comparing the solid and dashed lines of  Fig. \ref{relkin}, representing the carbon cross 
sections obtained from Eq. \eqref{sigma1} using relativistic and non relativistic kinematics, respectively. It clearly appears that in a kinematical setup corresponding to 
$|{\bf q}| \sim 585 \ {\rm MeV}$ at $\omega = \omega_{\rm QE}$ relativistic kinematics sizeably affects both position and width of the  quasi elastic peak. 

The factorisation {\em ansatz} of Eq. \eqref{fact1} can be readily extended to allow for a consistent treatment of the amplitudes 
involving one- and two-nucleon currents. The resulting expression is 
\beq
\label{fact2}
| N \rangle = | {\bf p} {\bf p}^\prime \rangle \otimes | m_{A-2} , {\bf p}_m \rangle \ ,
\eeq
where the states $| {\bf p} {\bf p}^\prime \rangle$ and $| m_{A-2} , {\bf p}_m \rangle$ describe two non interacting nucleons of momenta ${\bf p}$ and ${\bf p}^\prime$ and 
the $(A-2)$-particle residual system, respectively.

Using Eq. \eqref{fact2}, the nuclear matrix element of the two-nucleon current can be written in terms of  two-body matrix elements according to 
\begin{align}
\label{matel:2}
\langle N | j_{ij}^{\mu} | 0 \rangle   =
 \int d^3k d^3 k^\prime   M_m({\bf k},{\bf k}^\prime&)  
\langle {\bf p} {\bf p}^\prime | j_{ij}^\mu | {\bf k} {\bf k}^\prime \rangle  \ ,
\end{align}
with $M_m({\bf k},{\bf k}^\prime)$ given by
\begin{align}
\label{def:Mn}
M_m({\bf k},{\bf k}^\prime) = \left\{ \langle m_{(A-2)} , {\bf p}_m | \otimes \langle  {\bf k}  {\bf k}^\prime | \right\} | 0 \rangle \ .
\end{align}

From the above equations it follows that the evaluation of the nuclear transition matrix element involving the two-nucleon current reduces
to the calculations of the nuclear amplitude $M_m({\bf k},{\bf k}^\prime)$ and of the matrix element of the current operator between free nucleon states.
The former, being independent of momentum transfer, can be carried out using the non relativistic formalism, while the latter does not involve any 
approximations.

The connection with the spectral function formalism 
becomes apparent noting that the two-nucleon spectral function $P({\bf k},{\bf k}^\prime,E)$,
yielding the probability of removing {\em two nucleons} of momenta ${\bf k}$ and ${\bf k}^\prime$  from the nuclear ground state leaving 
the residual system with excitation energy $E$, is defined as [compare to Eq. \eqref{pke1}]~\cite{spec2}
\begin{align}
\label{def:pke2}
P({\bf k},{\bf k}^\prime,E) = \sum_m |M_m({\bf k},{\bf k}^\prime)|^2 \delta(E + E_0 - E_m) \ ,
\end{align}
with $M_m({\bf k},{\bf k}^\prime)$  given by Eq. \eqref{def:Mn}.

\begin{figure}[h!]
\includegraphics[scale=0.45]{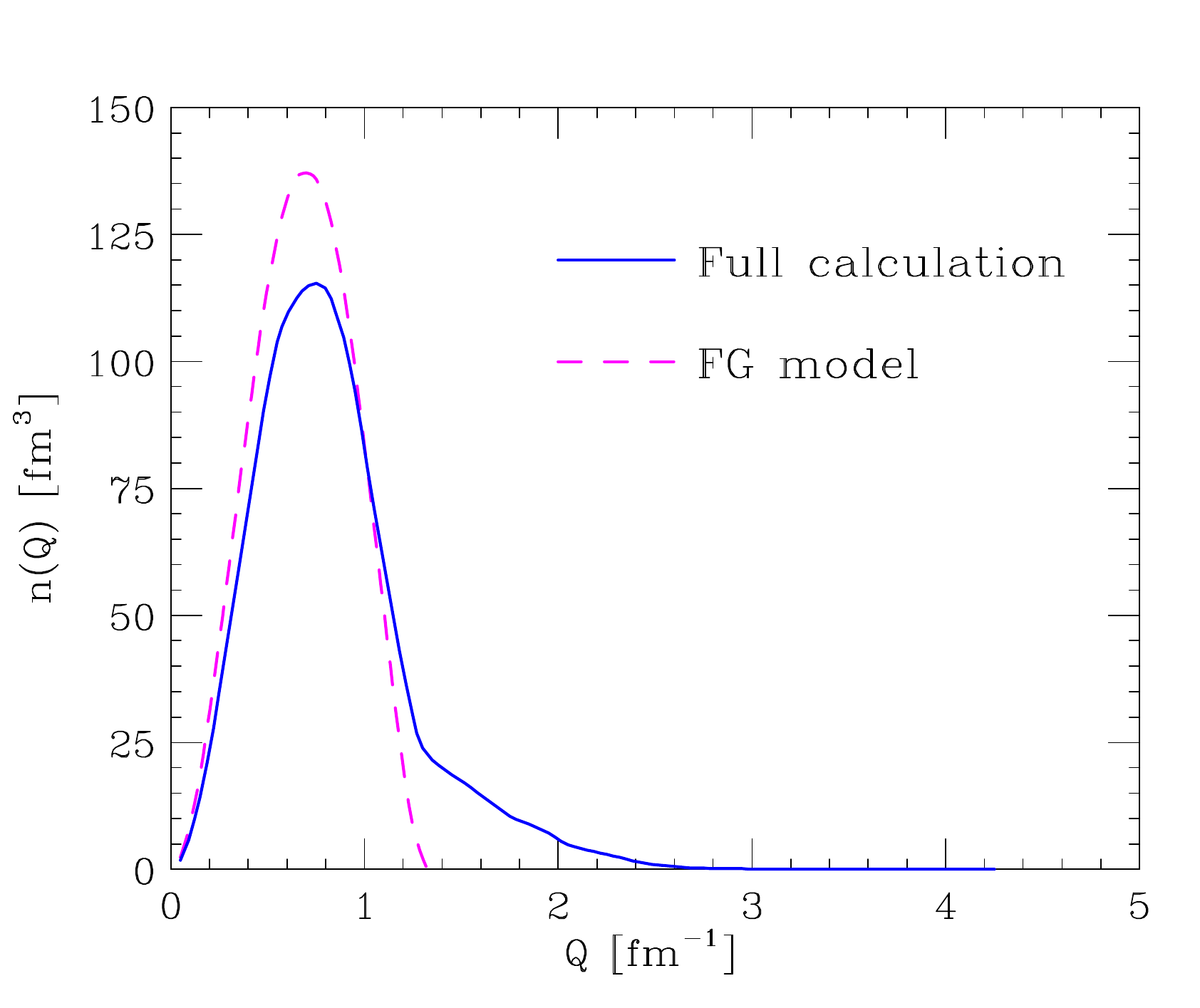}
\vspace*{-.1in}
\caption{(color online) Relative momentum distribution of a nucleon pair in isospin symmetric nuclear matter at equilibrium density.}
\label{SF}
\end{figure}

The two-nucleon spectral function of uniform and isospin symmetric nuclear matter at equilibrium density has been calculated 
 by the authors of Ref. \cite{spec2} using a realistic hamiltonian. The resulting relative momentum distribution, defined as
\begin{align}
\label{rel:dist}
n({\bf Q}) = 4 \pi |{\bf Q}|^2 \int d^3 K \ n\left( \frac{ {\bf K} }{2} + {\bf Q}, \frac{ {\bf K} }{2} - {\bf Q} \right)
\end{align}
where ${\bf K} = {\bf k}+{\bf k}^\prime$, ${\bf Q} = ({\bf k}-{\bf k}^\prime)/2$, and 
\begin{align}
n({\bf k},{\bf k}^\prime) = \int dE  \ P({\bf k},{\bf k}^\prime,E) \ ,
\end{align}
is shown by the solid line of Fig. \ref{SF}. Comparison with the prediction of the Fermi Gas (FG) model, represented by the dashed line,
 indicates that correlations give rise to a sizeable quenching of the peak of the distribution, along with  the appearance
 of a high momentum tail.
 
\subsection{1p1h and 2p2h contributions to the transition matrix element}
\label{rel2}

The extended factorisation {\em ansatz} 
discussed in the previous section  
provides a scheme allowing for a clear cut identification of the 1p1h and 2p2h contributions to the 
nuclear cross section in the presence of two-nucleon currents. 


Let us consider the contribution of 2p2h final states to the response tensor of Eq.~\eqref{response:tensor} 
\begin{widetext}
\begin{align}
\label{had2p2h}
W^{\mu\nu}_{2p2h} & = \int {d^3h\ d^3h^\prime\ d^3p\ d^3p^\prime} 
 \theta(k_F - |{\bf h}|) \theta(k_F-|{\bf h^\prime}| )  \theta(|{\bf p}| - k_F)\theta(|{\bf p^\prime}| - k_F)  \\
 \nonumber
 & \times \langle 0| J^\mu |{\bf h}{\bf h^\prime}{\bf p}{\bf p^\prime}\rangle \langle {\bf h}{\bf h^\prime}{\bf p}{\bf p^\prime}| J^\nu|0 \rangle 
 \delta(\omega + E_0 - E_{h h^\prime  p p^\prime})  \delta({\bf q} + {\bf h} + {\bf h}^\prime - {\bf p} - {\bf p}^\prime) \ , 
\end{align}
\end{widetext}
where ${\bf q}$ is the total momentum transfer, while ${\bf h}, {\bf h^\prime}$ and ${\bf p},{\bf p^\prime}$ are the momenta of the hole and particle states, respectively.
The structure of the current operator, involving one- and two-body terms, can be best understood from its momentum space expression
 \begin{align}
 \nonumber
J^\mu({\bf k_1},{\bf k_2}) & =  \int \,{d^3 x_1 d^3x_2} \  J^\mu({\bf x_1},{\bf x_2})  \  e^{-i ({\bf k_1}\cdot {\bf x_1} + {\bf k_2}\cdot {\bf x_2})}\\
&=  j^\mu_1({\bf k_1})  \delta({\bf k_2})+  j^\mu_2({\bf k_2})  \delta({\bf k_1})  + j_{12}^\mu({\bf k_1},{\bf k_2})  \ ,
 \end{align}
 showing how the  total momentum transfer, ${\bf q}~=~{\bf k_1}~+~{\bf k_2}$,  is shared between the two nucleons involved in the electromagnetic interaction, labeled by the indices $1$ and $2$.

Within the factorisation scheme, the matrix element of the one-nucleon current operator, can be readily evaluated  inserting  a complete set of states describing a non interacting nucleon. 
The resulting expression is 
\beq
\label{fact1b}
\langle 0| j^\mu_1 | {\bf h} {\bf h^\prime} {\bf p} {\bf p^\prime} \rangle= \int d^3 k\  \Phi^{h h^\prime p^\prime}_{k}    \langle{\bf k}|j^\mu_1 |{\bf p}\rangle , 
\eeq
with $\Phi^{h h^\prime p^\prime}_{k}$ defined by Eq. \eqref{def:phi2h1p}.

The calculation of the matrix element of the two-nucleon current exploits the fact that, in analogy with $P({\bf k},E)$, the two-nucleon spectral function of Eq. \eqref{def:pke2} can be separated into  two parts, characterised by their analytical structure.
The component corresponding  to {\em bound} 2h states of the $(A-2)$-nucleon system exhibits a pole located at $E=-(e_{k} + e_{k^\prime})$, whereas the
continuum states,  the dominant of which is the 3h1p state,  give rise to a smooth background. 

It follows that, within the factorisation scheme, the contribution to  $W^{\mu\nu}_{2p2h}$ arising from amplitudes involving {\em only} the two-nucleon current is obtained from the 
$2h$ component of $P({\bf k},{\bf k^\prime}, E)$,  which can be written in the form \cite{spec2}
\begin{align}
\label{2hsf}
P_{2h}({\bf k},{\bf k^\prime}, E) & = \int d^3h d^3h^\prime |\Phi^{h h^\prime}_{k k^\prime}|^2   \delta(E + e_{h}+ e_{h^\prime}) ,  \\
\nonumber
& \times \theta(k_F - |{\bf h}|)\theta(k_F - |{\bf h^\prime}|)  \ .
\end{align}
In the above equation, $\Phi^{h h^\prime}_{k k^\prime}$ is related to the overlap between the target ground state and the $2h$ state of the $(A-2)$-nucleon system through
\begin{align}
 \Phi^{h h^\prime}_{k k^\prime}   = \langle 0 | \{ | {\bf k} {\bf k}^\prime \rangle \otimes | {\bf h} {\bf h}^\prime   \rangle   \} \ .
\label{2hov0} 
\end{align} 
The diagrammatic analysis of the cluster expansion of $\Phi^{h h^\prime}_{k k^\prime}$ in uniform and isospin symmetric nuclear matter, carried out by the authors of Ref.~\cite{spec2}, 
shows that only unlinked graphs  (i.e.,  graphs in which the points reached by the ${\bf k}, \ {\bf k}^\prime$ lines are not connected to one other by any dynamical or statistical correlation lines) 
survive in the $A \to \infty$ limit, the contributions of linked diagrams being of order $1/A$.  It follows that
\begin{align}
\label{2hov}
\Phi^{h h^\prime}_{k k^\prime} & = \phi^{h}_{k} \phi^{h^\prime}_{k^\prime} \delta({\bf h}-{\bf k})\delta({\bf h^\prime} -{\bf k^\prime}) \ ,
\end{align}
where $\phi^{h}_{k}$ is the the Fourier transform of the overlap between the ground state and the $1h$ $(A-1)$-nucleon state, the calculation of which is discussed in Ref.~\cite{PKE}.

Collecting the above results, we can write the expression of the response tensor obtained from the extended factorisation {\em ansatz} as a sum of three contributions. The terms 
involving the squared amplitudes of the matrix elements involving one- and two-nucleon currents can be written in terms of the appropriate contributions to the 
one- and two-nucleon spectral functions, according to
\begin{widetext}
\begin{align} 
 W^{\mu\nu}_{2p2h,\rm{11}} = \int d^3 k \int dE P_{2h1p}({\bf k}, E) \langle {\bf k}|j^\mu_1|{\bf k+q}\rangle 
 \langle {\bf k+q}| j^\nu_1|{\bf k}\rangle\delta(\omega -E - e_{|k+q|})\theta(|{\bf k+q}| - k_F) 
 \end{align}
 and
 \begin{align}
 W^{\mu\nu}_{2p2h,\rm{22}} & = \int d^3 k d^3 k^\prime  d^3 p d^3 p^\prime\int dE P_{2h}({\bf k},{\bf k^\prime},E) 
 \langle {\bf k} {\bf k^\prime}| j^\mu_{12}| {\bf p}{\bf p^\prime}\rangle \langle {\bf p}{\bf p^\prime}| j^\nu_{12}| {\bf k} {\bf k^\prime}\rangle \\
\nonumber
&  \times \delta({\bf k}+{\bf k^\prime}+ {\bf q} - {\bf p} - {\bf p^\prime})\delta(\omega - E - e_{p}- e_{p^\prime})
\theta(|{\bf p}| - k_F)\theta(|{\bf p^\prime}| - k_F) \ .
 \end{align}
 \end{widetext}

The interference term, on the other hand, involves a product of the nuclear amplitudes entering the definition of the spectral functions.  The resulting expression is

 \begin{widetext}
\begin{align} 
\label{total}
 W^{\mu\nu}&_{2p2h,\rm{12}} = \int d^3 k\ d^3\xi\  d^3\xi^\prime\ d^3h\  d^3 h^\prime d^3p\ d^3p^\prime 
{\phi^{h}_{\xi}}^\ast {\phi^{h^\prime}_{\xi^\prime}}^\ast  \delta({\bf h}-{\bm \xi}) \delta({\bf h^\prime} -{\bm \xi^\prime})  \left[ 
 {\Phi^{h h^\prime p^\prime}_{k}}  \langle {\bf k}|j^\mu_1|{\bf p}\rangle + {\Phi^{h h^\prime p}_{k}}  \langle {\bf k}|j^\mu_2|{\bf p^\prime}\rangle  \right] \\
 \nonumber
&  \times  \langle {\bf p},{\bf p^\prime}| j^\nu_{12}| {\bm \xi}, {\bm \xi^\prime}\rangle  \  \delta({\bf h}+{\bf h^\prime}+ {\bf q} - {\bf p} - {\bf p^\prime})
\delta(\omega +e_{h}+ e_{h^\prime}-e_{p} - e_{p^\prime})
 \theta(|{\bf p}| - k_F)\theta(|{\bf p^\prime}| - k_F) + \rm{h.c.} \ .
\end{align}
\end{widetext}

Extensive numerical calculations of the electron-carbon cross section based on the formalism described in this paper 
are under way. However, they involve a number of non trivial novel developments, e.g. the derivation
of the two-hole contributions to the nuclear spectral function, the discussion of which will require a separate publication. 

\begin{figure}[h!]
\includegraphics[scale=0.45]{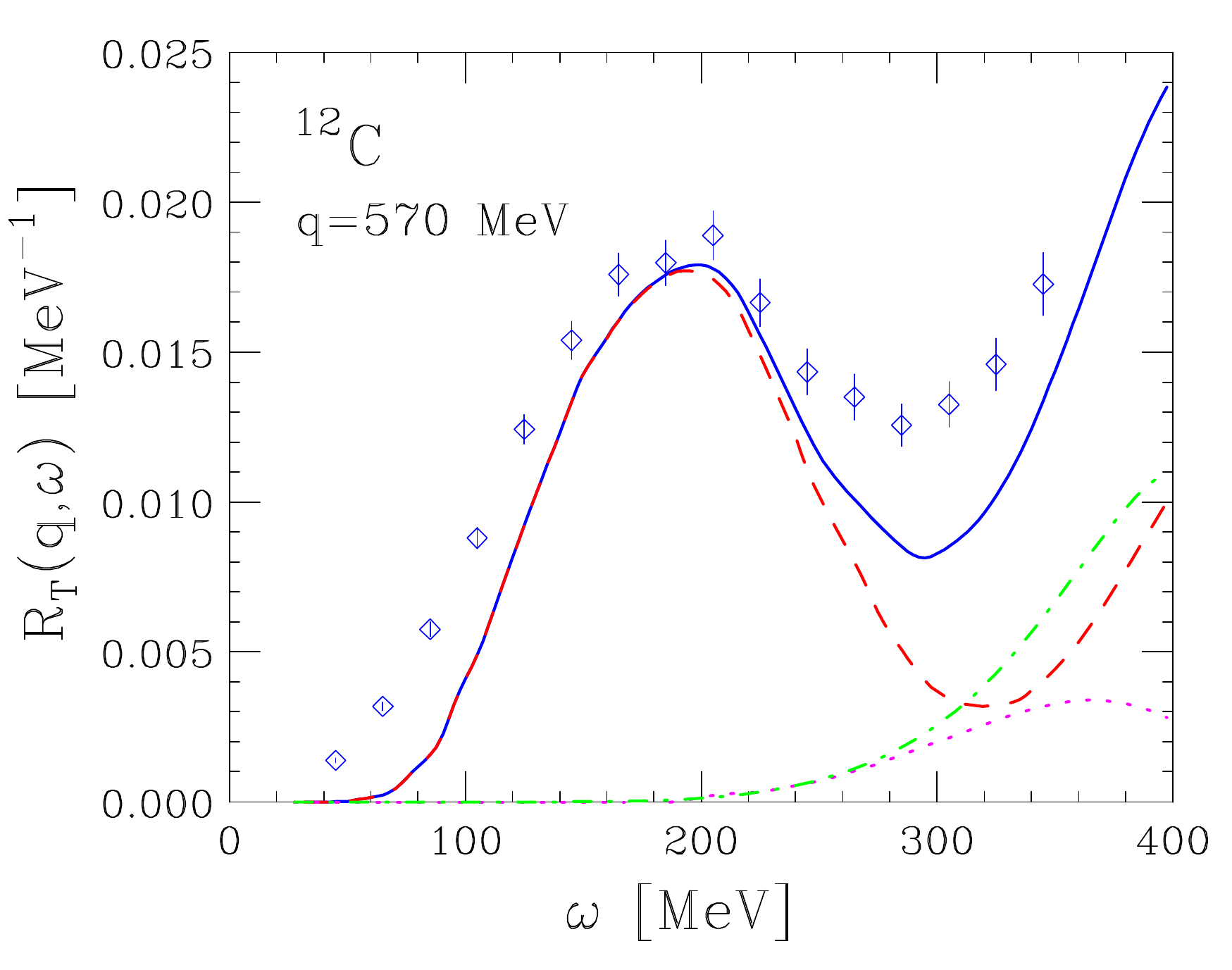}
\vspace*{-.1in}
\caption{(color online) Electromagnetic response of carbon in the transverse channel, at momentum transfer $|{\bf q}| = 570$ MeV.
The solid line represents the results of the full calculation, whereas the dashed line has been obtained including only amplitudes involving the 
one-body current. The contributions arising from the two-nucleon current are illustrated 
by  the dot-dash and dotted lines, corresponding to the pure two-body current transition probability and the interference term, respectively. 
The experimental data are taken from Ref.~\cite{jourdan}.}
\label{RT_12C}
\end{figure}

Figure \ref{RT_12C} shows the transverse electromagnetic response of carbon at $|{\bf q}| = 570$ MeV computed using the carbon spectral function of  Ref. \cite{LDA} and
approximating the two-hole spectral function of carbon with that of uniform nuclear matter, at density corresponding 
to Fermi momentum $k_F = 221$ MeV.  Note that this is {\em not} the same as working within the Fermi gas model. We 
use overlaps---the functions $\phi^{h}_{k}$ defined by Eqs.~\eqref{2hov0} and \eqref{2hov}---obtained within the {\em ab initio} approach of Refs.~\cite{PKE,BFF_split,spec2}, based on a realistic nuclear hamiltonian including two- and three- nucleon interactions. 

Owing to short range correlations, which move strength from the 1p1h to the 2p2h sector, the resulting  occupation of the momentum eigenstates is reduced by $\sim$ 20\%.

Interactions effects also affect the initial state energies of the knocked-out nucleons \cite{PKE,LDA}, thus shifting the threshold of 
the two-nucleon current contributions with respect to the predictions of the Frmi gas model \cite{dekker_MEC,depace_MEC}.

It has to be pointed out that the correlation contribution to the carbon spectral function of Ref.~\cite{LDA} is obtained from nuclear matter results. Therefore, 
the use of nuclear matter overlaps in the matrix elements of the two-nucleon current entering the interference terms appears to be consistent.

We have used the fully relativistic expression of the two-nucleon current described in Refs.~\cite{dekker_MEC,depace_MEC}, with the same form factors and 
$\Delta$-width.


The solid line of Fig.~\ref{RT_12C} represents the results of the full calculation, whereas the dashed line has been obtained including only the amplitudes involving the 
one-body current. The contributions arising from the two-nucleon current
are illustrated 
by  the dash-dot and dotted lines, corresponding to the pure two-body current transition probability and  the interference term, respectively. 
The latter turns out to be sizable, its contribution being comparable to the total two-body current response for $\omega \lsim 350$ MeV.
Although our results still need to be improved, and do not include the corrections taking into account the effects of FSI, in Fig. \ref{RT_12C} we have also included, for comparison, 
the data resulting from the analysis of Ref.~\cite{jourdan}.
 
\section{Conclusions}
\label{conclusions}

We have analysed the mechanisms---correlations in the initial and final states and coupling to meson exchange currents---leading to the excitation of 2p2h final states in the nuclear response 
to electromagnetic interactions.  In the non relativistic regime, in which highly accurate calculations consistently taking into account all these mechanisms are feasible, our results confirm 
the findings of Ref. \cite{Carlson:2002}. 

In the transverse channel, the contribution of processes involving the two-nucleon current is sizeable, and extends well into the kinematical region 
corresponding to energy transfer  $\omega \sim \omega_{\rm QE}$, in which single nucleon knock out is dominant. 

The important role played by interference between the amplitudes involving one- and two- body currents clearly implies that 
correlation effects must be included in any model aimed at  describing the nuclear cross section in the 2p2h sector. 
This point was clearly stated, over three decades ago, in the pioneering work of Ref. \cite{ref:Alberico_MEC}, the authors of which also remarked the 
inadequacy of the treatment of correlations based on lowest-order perturbative pion exchange.  
However, combining a realistic and consistent description of correlations and meson-exchange currents in the kinematical region in which the non 
relativistic approximation is no longer applicable involves serious difficulties.
 
To overcome this problem, we have developed a novel approach based on the  factorisation {\em ansatz} underlying the spectral function formalism,  widely and successfully employed 
to describe the nuclear response in the 1p1h sector. The preliminary results obtained within this approach, shown in Fig. \ref{RT_12C},  provide a fairly good description of the 
measured transverse response of carbon at $|{\bf q}| = 570$ MeV. 

A comparison between the results of Fig. \ref{RT_12C} and the GFMC results of Fig. \ref{4He} shows distinctive discrepancies in both magnitude and energy dependence of 
the two-body current contributions. While part of the disagreement is likely to originate from differences in the two-nucleon currents employed in Ref.~\cite{inversion}, as well as from the non relativistic 
nature of the GFMC calculations, the large interference contribution 
in the region of the quasi elastic peak observed in Fig. \ref{4He}  may arise from interference between amplitudes involving the one- and two-body currents and 1p1h final states. 
A careful analysis of these terms, that were found to be sizable in the pioneering work of Ref. \cite{Adelchi_MEC}, is being carried out, and will be discussed elsewhere.


The main assumption implied in the factorisation of the 2p2h final states is the treatment of the knocked out nucleons as free particles,  
which amounts to neglecting their interactions, both among themselves and with the spectator nucleons. Antisymmetrization under exchange
between any of the outgoing particles and the spectators is also disregarded.     

The factorized nuclear transition amplitudes involving the one-nucleon current can be corrected---to include the effects 
of final state interactions in the quasi elastic sector---using an extension of the spectral function formalism, as discussed in Ref.~\cite{benharFSI}.
The resulting modifications lead to (i) a shift in energy transfer of the differential cross section, arising from interactions 
between the knocked out nucleon and the mean field of the recoiling 
nucleus, and (ii) a redistribution of the strength from the quasi free 
peak to the tails, resulting from rescattering processes. Theoretical studies of 
electron-nucleus scattering suggest that in the kinematical region 
relevant to the MiniBooNE analysis the former mechanism---which does 
not involve the appearance of 2p2h final states---provides the dominant 
contribution, and can be taken into account through an optical potential  \cite{uncertainty}. 

The corrections to the factorized amplitudes involving the two-nucleon current also include
interactions between the two knocked out particles. A careful analysis of these 
processes is certainly needed. However, the results of Shen et al, who carried 
out an accurate calculation of the neutrino-deuteron
cross section over a broad kinematical range, suggests that their  effect becomes negligibly small at beam 
energies larger than $\sim 500$ MeV \cite{shen}.

In conclusion, we believe that the approach described in this paper provide a viable and promising scheme for the 
development of a unified treatment of processes involving one- and two-nucleon 
currents, applicable in the kinematical region relevant to neutrino oscillation searches. 
Therefore, it may in fact be regarded as a step towards the  {\em new paradigm} advocated 
by the authors of Refs. \cite{coletti,benhar:neutrino2010}.

We are deeply indebted to A.M. Ankowski, J. Carlson, S. Gandolfi, C. Mariani, S. Pieper and R. Schiavilla for many 
illuminating discussions. The work of OB and NR was supported by INFN under grant MANYBODY. 
 The work of AL was supported by the U.S. Department of Energy, Office of Science, Office of Nuclear Physics, under contracts DE-AC02- 06CH11357.
 Under an award of computer time provided by the INCITE program, 
this research used resources of the Argonne Leadership Computing
Facility at Argonne National Laboratory, which is supported by the
Office of Science of the U.S. Department of Energy under contract
DE-AC02-06CH11357.

\end{document}